# Detecting Synthetic Phenomenology in a Contained Artificial General Intelligence

Jason M. Pittman     Ashlyn Hanks

jpittman, ahanks {@highpoint.edu}
High Point University
High Point, NC 27268

*Abstract* -- Human-like intelligence in a *machine* is a contentious subject. Whether mankind should or should not pursue the creation of artificial general intelligence is hotly debated. As well, researchers have aligned in opposing factions according to whether mankind can create it. For our purposes, we assume mankind can and will do so. Thus, it becomes necessary to contemplate how to do so in a safe and trusted manner- enter the idea of *boxing* or *containment*. As part of such thinking, we wonder how a phenomenology might be detected given the operational constraints imposed by any potential containment system. Accordingly, this work provides an analysis of existing measures of phenomenology through qualia and extends those ideas into the context of a contained artificial general intelligence.

## 1. Introduction

True artificial intelligence (AI), often referred to as artificial general intelligence (AGI) or machine consciousness (MC) is a long-held goal. However, for the many that claim development of an AGI is inevitable, there are just as numerous asserting AGI is an unreachable, unattainable end state. Despite the eventual truth outcome of those propositions, the literature suggests that concurrently developing a means of containing any such machine superintelligence is a worthwhile endeavor. However, the field of artificial agent containment is quite young and still developing a stable foundation.

Accordingly, the problem we sought to address is the lack of defined synthetic phenomenology detection criteria for contained artificial agent. We elaborate on our journey by exploring potential answers to the question of *how we might detect synthetic phenomenology in a contained artificial agent*. While the reach of this work is speculative, we intend to explore a possible answer to this problem by critically reviewing existing literature and constructing a detection criteria matrix extension to the ontology for artificial agent containment.

The remainder of this work is organized as follows: section two presents a synthesis of germane literature; section three reveals the assumptions we have found in the background material as well as ideas for addressing such; and finally, section four describes our conclusions and recommendations.

## 2. Background

Fleshing out the relevant background literature before delving into a potential means of detecting synthetic phenomenology in a contained AI will let us properly situate associated concepts. The goal is not to provide an exhaustive review but simply to position critical and relevant research in an enveloping context. Thus, given the topic at hand, the first step in our background will be to explore *qualia* both in a conceptual light as well as specifically in the context of AI. This is followed by an exploration of the existing proposals for detecting consciousness, qualia, and phenomenological experiences. Lastly, we address the ideas of *synthetic phenomenology* and *containment*.

## 2.1 Qualia

Qualia are understood to be the way something (external, reference our mind) appears to us (Arrabales et al., 2010). In this manner, we can take qualia to be the subjective experience of a sight, a scent or a sound; essentially, any sense-based interaction with our environment and the objects within it. Of course, while there is debate regarding the nature of qualia or even if qualia are valid mental constructs (Dennett, 1991), there is also agreement that qualia are coupled to consciousness. Thus, we are left with an open question of whether qualia are exclusively bound to natural minds. There could be, after all, some means to establish artificial qualia.

Towards that end, Arrabales et al. (2010) argue that developing engineering principles (perception, introspection, and reportability) for synthetic phenomenology is a precursor to constructing a formal definition for artificial qualia. A subtle implication is that artificial qualia may not be equivalent to natural qualia in effect if not also in process. That is, inquiring about qualia through a variety of engineering principles may provide insight into designing computational access to qualia as well as computational generation of qualia. Further, such engineering principles may lend insight into how to test for qualia and phenomenological events.

## 2.2 Tests for Qualia and Phenomenology

In fact, the literature contains a wealth of proposed tests for qualia and phenomenology. However, there is a stark difference between testing *for* consciousness and testing the *degree of* consciousness. Unfortunately, the literature does not seem to contain any significant analysis of these proposed tests both in terms of relationships between efforts and regarding *prima facie* validity. We leave such questions for future work in favor of establishing a cataloging of work.

Elamrani and Yampolskiy (2019) review a battery of tests targeting machine consciousness. The tests are documented in chronological order by creation date and provide a brief description. Of note, the list of tests begins in 1950 with the *Turing Test* and extends through 2014 with the *Human-Like Conscious Access*. The work then compares the tests based on operational criteria while organized the tests into two categories – architecture-based tests and behavior-based tests. Further, there is extensive notation related to what fields are proposing the tests and how impactful each are by totaling the citations on Google Scholar. Perhaps unsurprisingly, the Turing Test is by far the most common, with only three other tests reaching above 100 citations – Phi, Total Turing Test, and 5 Axioms.

Still, Arrabales et al. (2009) point out the metrics for artificial consciousness is an open problem and requires more research. Furthermore, Arrabales et al. suggest measures must use third person which is similar in concept to Dennett's (1991) argument. Based on that, the implication is that measurements in the same vein as the Turing Test are inherently flawed because such instruments measure from a first-person lends. Instead, according to Arrabales et al., measures ought to legitimately assess whether cognition is present; that is, the measures must apply to any machine consciousness, be problem domain independent, computable in reasonable time, and provide multidimensional characterization.

On the topic of the Turing Test, Schweizer (2012) examined the possibility of a Turing Test aimed at measuring whether a *computational artefact* is the subject of conscious experience. Here, Schweizer argued the Turing Test is too weak to detect such artefacts. As a proposed remedy, Schweizer proposed a Q3T (Qualia Total Turing Test). Rather than testing for qualia through first-person questioning, the Q3T would be designed as a robust behavioral test. This of course has limitations as well since the concept presupposes an ability to behave as well as display behavior. Likewise, this presupposes that the interlocutor can perceive such behavior.

Yampolskiy (2017) also proposed a derivation of the Turing Test. Here, the underlying rationale is that since humans solve CAPTCHA problems using *a posteriori* reasoning (i.e., stored qualia), then any artificial intelligence also capable of doing so must possess some phenomenological capability. Further, unlike other tests for consciousness which assert a binary state (conscious versus non-conscious), Yampolskiy's idea allows for discrete probabilities of phenomenology. Thus, agents that have not experienced enough conditions present in the CAPTCHA are not necessarily precluded from detecting a solution. In fact, the agent does not need *identical* qualia to be successful.

*2.3 Synthetic Phenomenology*
Given the limitations and shortcomings of existing tests for consciousness, research has turned to potential alternative detection characteristics. Put simply, if consciousness itself is not directly detectable, perhaps it may be possible ascertain if a simulated consciousness has phenomenology.

Along those lines, Gamez (2005) attempted to address whether it might be possible for a *machine* to possess mental states (i.e., qualia). The solution, according to Gamez, resides in detecting the characteristics or conditions most often associated with consciousness. The problem of course is that existing instrumentation simply is not powerful enough, finely tuned to the degree necessary to measure phenomenology at a physical level. This is where synthetic phenomenology may yield some insight.

Synthetic phenomenology introduces the notion that phenomenal states can be modeled by, or *within*, an artifact (Chrisley, 2009). Abstractly, this artifact could be any object in physical space. Concretely, in the context of containment, we are implying computational machines (i.e., computers, robots, and the like). Intentionally, if any such artifact is capable of phenomenal states, identifying mental content (qualia) of the artifact as well as analyzing the structure of such mental content may be possible. Implicitly, this means we may be able to identify discrete parts that are phenomenally conscious.

Fortunately, Gamez (2005) proposed a scale to measure discrete parts in a probability distribution of machine consciousness based on the idea of synthetic phenomenology. Of course, this assumes synthetic phenomenal states are at least partially representational of non-synthetic (human) states. Further, this assumes that both the artificial agent or artifact are reachable by any such instrumentation and that the associated synthetic qualia are observable. Such assumption is likely to become amplified as a difficult challenge given the need to explore artificial superintelligence in some contained substrate.

*2.4 Containment*
Concurrent to the notions of artificial agents and artificial qualia or phenomenology is the idea that such an artificial agent ought to be *contained* or *boxed*. As of now, there are no practical containment solutions. However, the concept has received increasingly attention in the research community. Clearly the potential of an artificial agent behaving - unintentionally or deliberately- in a manner contrary to mankind's best interests is something that ought to be seriously considered. Moreover, the problems associated with *how* to contain or box an artificial agent are quite complex. In the context of this work, we consider the concept of containment as a compounding factor for potential qualia or synthetic phenomenology testing issues.

To explain further, Babcock et al.(2016), Sotala and Yampolskiy (2015), and Babcock et al. (2017) suggested that artificial agent come to be functionally within a containment system which can prevent, detect, or otherwise control the ingress and egress of interactions between an artificial agent and humans. Conceptually though, both artificial general intelligence and the containment of such entities are demanding abstractions. This notion is not surprising given that the nature of autonomy itself is hidden from our understanding and instrumentation (Wiese, 2018). Further confounding variables exist because the heart of the containment problem is essentially a cybersecurity concern, yet no viable branch of cybersecurity is positioned to address containment.

Overall, containment must assert *control* in areas where artificial agent might maliciously or irresponsibly interact with humans (Babcock et al., 2016). Such controls are predicated on the assumption that artificial agent not only possesses a catastrophic, even existential, risk to human society but that it will act on such risk to further its own existence (Muller, 2014; Sotala & Yampolskiy,2014; Chalmers, Awret, & Appleyard, 2016; Amodei et al., 2016; Yudkowsky, 2008). However, by asserting any such control, containment establishes a barrier between artificial agent and the instrumentation seeking to assess consciousness. Consequently, a serious issue is the degree to which the containment barrier is permeable.

Bishop (2018) pointed out a significant problem in that we are left to identify, *a priori*, what knowledge is necessarily restricted by containment. Here, *a priori* reasoning is seemingly required because external entities likely will not be able to interact directly with the contained intelligence while, concurrently, the contained intelligence will not be able to access information about anything external to the containment. The resulting challenges created by containment relative to testing for and testing the degree of synthetic phenomenology cannot be overstated. In some fashion, without viable instrumentation and testing profiles, we are screaming into a void.

With that said, Bishop (2018) proposed an *artificial consciousness test* (ACT) for contained artificial intelligences. Subsequently, Kelley (2019) suggested a framework for testing *mediated* super intelligent artificial agents. As of this writing, these works represent the extent of the investigation.

**3. Discussion**
To be certain, there is a robust and lively literature supporting qualia, tests for qualia, synthetic phenomenology, and AI containment. The purpose of our background was to present summary of what we perceive to directly contribute to the discussion of how synthetic phenomenology or synthetic qualia might be observed in contained AI. A critical portion of such discussion must account for assumptions and limitations. After all, only after outlining where some arguments may be insufficient or weak can we develop reasonable detection criteria for contained AI.

In our estimation, existing work relies on four critical assumptions. While such assumptions may or may not prove to be valid, the quality of future work can benefit from understanding these implicit propositions. Additionally, assumptions laid bare impart transparency to how we implement synthetic phenomenology detection in contained artificial agents may work.

One assumption is that containment is an on-off, zero-sum calculus. In other words, artificial agent is either contained or not contained with the former representing a win-lose for humans to artificial agent and lose-win when artificial agent is not contained. This position certainly is speculative in that the relationship between humans and artificial agent may be nonzero sum at all or containment may neutralize any zero-sum tendencies.

Further, existing work assumes that MC will manifest in a manner consistent with detection from an anthropomorphic lens. On one hand, such an assumption is rational insofar as the singular form of consciousness we know of seems to be a universal consciousness. On the other hand, the universality of consciousness as we perceive it may be the result of a singularity in our own consciousness that renders us blind to other instantiations.

As well, existing work assumes that any observed response to synthetic phenomenology detection is not the result of illusion. Here, we ought to exercise caution in aligning *illusion* with the appropriate side of the detection equation: observer or observed. Assuming illusion is not a potential factor, even if only by omission, may be rationale for the observer side because of the capacity to control biases (i.e., operationalized illusion) in the detection instrumentation. Yet, assuming that an observed artificial agent is not capable of operating under the illusion of consciousness may be a limiting factor in detecting synthetic phenomenology.

Finally, much of our work assumes any future containment system for artificial agent is time based and thus permeable to stimuli as well as responses. A temporal containment system is not an obvious violation of known universal constraints but also is entirely theoretical. As companion fields of inquiry (e.g., physics, computer science, and so forth) expand their knowledge, the validity for this assumption will become clearer. Moreover, we assume that information passed back and forth across a time containment barrier will retain expected integrity and confidentiality values. We know these values are consistent in human to human interaction *across* time but not if the same holds true for human to artificial agent interaction *through* time.

## 4. Conclusion

Time in fact may present one possible answer to the problem. One of us suggested *time* as an idea for how such containment may be manifestable (Pittman, n.d.) in a fashion that assures containment without impeding interaction with contained artificial agents. The notion involves embedding the artificial agent within some form of time dilatation. This would need to be done at a germative stage of development, prior to the agent reaching sufficient phenomenological capacity to perceive the environment around it. Thereafter, the agent is rendered permanently *safe* by virtue of existing in a fixed past time relative to our present time.

Overall, we are left to *speculate* about what criteria may be necessary to detect the presence of a synthetic phenomenology in a contained artificial agent. What is known is tests for contained agent synthetic phenomenology would need to be created that would probe agents for characteristics of qualia. However, the issue with that is there is no way for certain to know if something is truly conscious and having experiences based solely on our current understanding of human analogues.

While scholars such as Searle (1980), Dennett (1991) and Chalmers (1993; 2018) have written extensively about the problem of detecting consciousness, little work has been done with *contained* artificial agents. As well, while a robust literature exists for measuring intelligence based on Turing's work, detecting consciousness, and furthermore self-consciousness, is not synonymous with detecting intelligence.

In conclusion, developing a method to detect synthetic phenomenology in a contained artificial agent also has significance for detecting such in non-contained artificial agent a narrow AI, as well as potentially for forms of synthetic or non-human consciousness. Future work should continue along the line of inquiry established by Bishop (2018) and Kelley (2019). Moreover, as research into containment continues to build forward, the community ought to periodically revisit old ideas with an eye towards reapplying instrumentation that may become viable due to underlying technological changes.